\documentstyle[twoside,fleqn,espcrc2]{article}

\def\ApJ{{\it Astrophys. J.} }
\def\ApJL{{\it Astrophys. J. Letters} }
\def\AA{{\it Astron. \& Astroph.} }
\def\ApP{{\it Astroparticle Physics} }

\def\PRL{{\it Phys. Rev. Letters} }

\def\MNRAS{{\it Month. Not. Roy. Astr. Soc.} }
\def\PASP{{\it Publ. Astron. Soc. Pacific} }
\def\AJ{{\it Astron. J.} }

\def\etal{{\it et al.} }
\def\simle{\lower 2pt \hbox {$\buildrel < \over {\scriptstyle \sim }$}}
\def\simge{\lower 2pt \hbox {$\buildrel > \over {\scriptstyle \sim }$}}
\newif\iffigures
        \figuresfalse 

\newcommand{\AmS}{{\protect\the\textfont2
  A\kern-.1667em\lower.5ex\hbox{M}\kern-.125emS}}

\title{Origin of the highest energy cosmic rays}

\author{Peter L. Biermann\address{Max Planck Institut f{\"u}r Radioastronomie,
Bonn, Germany},
Eun-Joo Ahn\address{Department of Physics,
Seoul National University, Seoul, Korea},
Gustavo Medina-Tanco\address{Institute of Astronomy and Geophysics,
University of Sao Paulo, Sao Paulo, Brasil},
and Todor Stanev\address{Bartol Research Institute,
University of Delaware, Newark, DE, USA}}

\begin{document}

\begin{abstract}
Introducing a simple Galactic wind model patterned after the
solar wind we show that back-tracing the orbits of the highest energy cosmic
events suggests that they may all come from the Virgo cluster, and so probably
from the active radio galaxy M87.  This confirms a long standing
expectation. 
Those powerful radio
galaxies that have their relativistic jets stuck in the interstellar
medium of
the host galaxy, such as 3C147, will then enable us to derive limits on the
production of any new kind of particle, expected in some extensions of the
standard model in particle physics.  New data from HIRES will be crucial in
testing the model proposed here. 
\vspace{1pc}
\end{abstract}

\maketitle

\section{Introduction}

 The origin of the highest energy particles observed in the universe
 continues to present a major challenge to physicists.  These particles
 reach energies as high as $3 \, 10^{20}$ eV.  If protons,
their flux is expected to drop sharply at $5 \, 10^{19}$ eV due to
the interaction with the microwave background, commonly referred to as the
 Greisen-Zatsepin-Kuzmin- or GZK-cutoff after its discoverers.  However, the
number of
 particles known to be beyond $10^{20}$ eV continues to increase, with
now 14
 published \cite{LB99,AGASA98,AGASA99}, and a further 10 expected from new
observations with  HIRES and a reanalysis of the Yakutsk 
data.  

 There are three basic difficulties:  

 First, we need to find a site
 that can produce particles at such energies
 \cite{Hillas84,JGCR}.  There is only one class that has
 been argued to require protons at such energies in the source, namely radio
 galaxies with powerful jets and/or hot spots, such as M87 \cite{BS87}
 or Cyg A.  M87 has been under suspicion to be the primary source for
 ultra high energy cosmic rays for a long time, e.g.,
\cite{GS63,HP80}.
 Gamma ray bursts are another possible class.

Secondly, there is the additional difficulty of getting these
particles
 to us, that is, overcoming the losses in the bath of the cosmological
 microwave background.  That implies that the
source should not be
 very much further than 20 Mpc.

The third difficulty is the explanation of the nearly isotropic
 distribution of the arrival directions of these events.

 There are several ways out of these difficulties.

 Here
 we present the consequences of introducing a
 magnetic Galactic wind in analogy to the solar wind.
 The magnetic field of the wind bends the
 particle orbits without adding substantial travel time.

\section{A model for a magnetic galactic wind}

 It has long been expected that our Galaxy has a 
wind, e.g.,\cite{Burke68}
 akin to the solar wind~\cite{Parker58}.  Recent 
modelling~\cite{Galwind99}
 shows that such winds can be quite fast, and ubiquituous.

 It seems plausible that this wind is powered by the combined action
 of cosmic rays and magnetic fields
 and starts in the hot phase of the interstellar medium
 seen in X-rays by ROSAT \cite{ROSAT97}.  

 Parker has shown~\cite{Parker58} that in a spherical wind the azimuthal
 part of the magnetic field  quickly becomes dominant with
 $B_{\phi} \, \sim \, \sin \theta /r$ in polar coordinates.

 Cosmic ray driving is similar to radiation driving of winds in massive
 stars, and so \cite{SB97} magnetic fields can lead to an increase of
 the momentum of the wind.
 
 The data on the sign of the azimuthal component show that in the part
of the
sky most relevant for calculating
 orbits of energetic charged particles in our Galaxy, the field points
to the
 direction of galactic longitude about 90 degrees
 \cite{Simard,Kronberg94}.  That
 means immediately that positively charged particles traced
 backwards have their origin above us, at high positive galactic latitudes.

  We adopt the simplest possible model.  We assume
 that the magnetic field in the galactic wind has a dominant azimuthal
 component, and ignore all other components.  We assume that this azimuthal
 component has the same sign everywhere \cite{Krause98}.  
 Most measures of magnetic field underestimate its strength.
 Therefore we will consider for reference a model which has a field
 strength near the Sun of 7 microGauss; this is close to the
best estimates for the total local field~\cite{Beck96}.  The second
 parameter, the distance to which this wind extends, is more uncertain:
 Our Galaxy dominates its near environment well past our neighbor, M31,
 the Andromeda galaxy, and might well extend its sphere of influence to
 half  way to M81.  Therefore we will adopt as outer the halo wind radius
 half the distance to M81, 1.5 Mpc.

\section{Tracing the path backwards}

 To follow the particle trajectories in the Galactic halo we trace
 protons 
backwards, e.g., \cite{Stanev97,GMT98c,Sigl98b}
 from their arrival direction at Earth.
 We use the 14 published cosmic ray events above 10$^{20}$ eV,
 the list from Watson (included in \cite{LB99}) and the new list
 from AGASA \cite{AGASA99}. There is a big uncertainty with
 the energy estimate of the highest energy
 Yakutsk~\cite{Yakutsk99a} event,
 which we therefore exclude from the present analysis, and hence
 we arrive at a final tally of 13 events used.

 The interesting result of these model calculations is
 that the directions of all tracks point North.
All events are consistent with arising originally from
 Virgo A. Since these particles are assumed to be accelerated
 out of cosmic gas, about 1/10 of all particles may be
 Helium nuclei with the same energy per particle.
 If the two highest energy events are in fact He nuclei, all 13 events
 point within 20 degrees of Virgo A.

 If Virgo A is indeed the acceleration site of the highest energy
 cosmic ray events, they all require additional systematic bending at a
ten to
 twenty degree level. Such bending could be easily accomodated within
 the plausible magnetic field strength within the supergalactic sheet
 from here to Virgo~\cite{Vallee,Ryu98}.

How critical are our assumptions for these results?

The assumption of the symmetry of the magnetic field
 above and below the Galactic disk is important.

The value of the magnetic field, here adopted as
 7 microGauss, for the wind near the Sun, is a key parameter.
If the magnetic
 field were considerably weaker, the focussing
 would be largely removed.  

The scale of the Galactic wind here 1.5 Mpc, is not
 a critical parameter, since the calculations show that most of the
 bending happens within the first few 100 kpc.

\section{Discussion and implications}

This particular model can be tested in several ways; some of these tests are
direct tests with data, and others take the form of a consistency check:

The most important test is obviously with more data, and the HIRES
data are soon to be released and will allow straight tests to be made.  For
instance, the apparent focussing is fully maintained at $5 \; 10^{19}$ eV.

The very concept of a magnetic wind, driven by cosmic rays,
 but with an initial magnetic field as strong as that in the disk,
 needs to be examined more closely.

There is one important consequence:  If the model proposed here could be
confirmed, then it would constitute strong evidence that all powerful
radiogalaxies produce high energy cosmic rays, and that they do
this at a good fraction of their total power
output.  This then implies that compact radio galaxies \cite{ODea98,FB98}
do provide a good test bed for particle interactions,
 since they have a large
screen of interstellar gas around the radio hot spots and jets as seen in
mm-wavelength radio data \cite{Papa99}.
  There the paradigm is materialized of
having a gigantic accelerator, and a beam dump.  These radio galaxies
may be
used for particle interaction experiments in the sky. If a significant
 correlation in arrival direction between ultra high energy cosmic rays
 and this specific class of radio quasars could be confirmed~\cite{FB98},
 then properties of new particles could be constrained. 

 In summary, we propose here that a very simple model for a
 Galactic wind rather analoguous to the Solar wind, may allow
 particle orbits at $10^{20}$ eV to be bent sufficiently to
 allow ``super-GZK'' particles to get here from M87,
 and also explain the apparent isotropy in arrival directions.

 {\bf Acknowledgments.}
This work was started at the 1999 astrophysics summer school
at the Vatican Observatory; both E.-J.A. and P.L.B. are
grateful to George Coyne, S.J. and his colleagues for inviting us and for
the fruitful and inspiring atmosphere at these schools.
 P.L.B, G.M.-T. and T.St. would
like to acknowledge also useful discussions with Rainer Beck, Venya Berezinsky,
 Pasquale Blasi,
 E. Boldt, Glennys Farrar,
 J.L. Han, Anatoly Ivanov, Randy Jokipii, Frank Jones, Hyesung Kang,
 Phil Kronberg, Martin
Lemoine, Friedrich Meyer, Hinrich Meyer, Biman Nath,
Dongsu Ryu, Michael Salomon, G{\"u}nther Sigl, Alan
Watson, Arnold Wolfendale and many others.
 We thank Phil Kronberg for a careful reading of an earlier manuscript.
Work with P.L.B. is partially supported by a DESY-grant,
G.M.-T. is partially supported by the Brazilian agencies
FAPESP and CNPq, T.St. is supported by NASA grant NAG5--7009,
 and PLB and TST have jointly a grant from NATO.

\end{document}